\RequirePackage{fix-cm}
\documentclass{svjour3}                     
\smartqed  
\usepackage{graphicx}
\usepackage{bbding}
\usepackage{upgreek}
\usepackage{url}
\usepackage{color,xcolor}
\usepackage{amsmath}
\usepackage{bm}
\usepackage[colorlinks,
            linkcolor=blue,
            anchorcolor=blue,
            citecolor=blue,
            urlcolor=blue,
            ]{hyperref}
\usepackage{mathptmx}      
\usepackage{titletoc}
\usepackage{amssymb}
\usepackage{flushend}

\newcommand{\av}[1]{\left<#1\right>}

\newcommand{\br}[1]{\left( #1 \right)}
\newcommand{\sbr}[1]{\left[ #1 \right]}
\newcommand{\lbr}[1]{\left\{ #1 \right\}}

\newcommand{\bx}{\boldsymbol{x}}

\newcommand{\pa}[1]{\partial_{#1}}

\newcommand{\ovl}[1]{\overline{#1}}

\newcommand{\dd}{\mathrm{d}}

\emergencystretch=\maxdimen
\titlecontents{section}[0pt]{\addvspace{2pt}\filright}              {\contentspush{\thecontentslabel\ }}              {}{\titlerule*[8pt]{.}\contentspage}

\hypersetup{
    bookmarks=false,         
    colorlinks=true,       
    linkcolor=blue,          
    linkbordercolor={1 1 1}
}

\journalname{Acta Mech. Sin.}
\begin{document}
\rmfamily
\title{Global expressions for high-order structure functions in Burgers turbulence
}
\subtitle{({\it Acta Mechanica Sinica})}

\titlerunning{Global expressions for high-order structure functions in Burgers turbulence}        

\author{          Jin-Han Xie
       }


\institute{{\Envelope} Jin-Han Xie \at
            \email{jinhanxie@pku.edu.cn} \at \at
		    Department of Mechanics and Engineering Science at College of Engineering and State key laboratory for turbulence and complex systems, Peking University, Beijing, 100871, PR China
           }
\date{\copyright {\it Acta Mechanica Sinica}, The Chinese Society of Theoretical and Applied Mechanics (CSTAM) 2020
}


\maketitle


\begin{abstract}
Since the famous work by Kolmogorov on incompressible turbulence, the structure-function theory has been a key foundation of modern turbulence study.
Due to the simplicity of Burgers turbulence, structure functions are calculated to arbitrary orders, which provides numerous implications for other compressible turbulent systems. 
We present the derivation of exact forcing-scale resolving expressions for high-order structure functions of the burgers turbulence.
Compared with the previous theories where the structure functions are calculated in the inertial range based on the statistics of shocks, our expressions link high-order structure functions in different orders without extra information on the flow structure and are valid beyond the inertial range, therefore they are easily checked by numerical simulations.
\keywords{Structure function \and Burgers turbulence \and turbulence theory}
\end{abstract}
\vspace{1 cm}

\section{Introduction}

Burgers equation \cite{Burgers1948} is almost the simplest model to study the shock-involved compressible turbulence \cite{Landau1987}. 
Similar to the vortex turbulence, e.g. \cite{Kolmogorov1941}, the study of Burgers turbulence focuses on the inertial range where the forcing and dissipation are believed to be irrelevant.
But due to the simplicity of Burgers equation, more information such as the probability distribution of the velocity is calculated \cite{Bouchaud1995}.

Structure functions are used to quantify the statistical properties of turbulence. 
In vortex-dominant turbulence, the exact theory for structure functions is most explored for the third order in the inertial ranges when the flow is incompressible, e.g., three-dimensional (3D) turbulence \cite{Kolmogorov1941}, two-dimensional (2D) turbulence \cite{Lindborg1999,Bernard1999,Yakhot1999}), turbulence with bidirectional energy transfer \cite{Alexakis2018,Xie2019,Xie2019b} and anisotropic sheared turbulence \cite{Casciola2003,Wan2009,Wan2010}.
The third-order structure function is important as it provides information on energy transfer across scales.
But to fully understand the statistics high-order structure function is also need, however, for the high-order structure functions the number of unknown independent variables is larger than the number of equations \cite{Hill2001}, leaving the expressions for high-order structure functions not directly calculated.
The situation becomes more complicated when the flow is compressible (cf. \cite{Galtier2011,Wang2013,Banerjee2014,Chen2015,Sun2017}), where the steady correlation equations contain not only velocity structure functions but also the density and pressure.
Even for the low-order structure functions corresponding to the energy transfer across scale, compressible turbulence is still a complicated target because of the coexistence of multiple modes with distinctive features \cite{Chen2021}. 
So to obtain a partial understanding of high-order structure functions in compressible turbulence this paper focuses on the less complicated Burger turbulence.

In the inertial range, \cite{E1999} made impressive progress by deriving the expressions of structure functions for Burgers equation in all orders. 
Their asymptotic derivation bases on the shock statistics such as the density and moments of the shocks, which makes it not easily justified by numerical simulations.

This paper derives the exact relations of high-order structure functions for Burgers equation following the procedure based on the K\'arm\'an-Howarth-Monin (KHM) equation \cite{Monin1975,Frisch1995}. 
So we do not need to consider the flow structures such as the shocks, therefore, our expressions avoid the appearance of shock statistics and are easier to be justified by numerical simulations. 
Following the KHM approach, odd-order structure functions are obtained in the inertial range \cite{Cardy2008,Falkovich2011}.
Comparing with these work, we capture the forcing (energy injection) scale beyond the inertial range (cf. \cite{Xie2018,Xie2019,Xie2019b}), which also makes our expressions easier to be justified by numerical results since identifying the inertial range potentially brings about artificial error.
Also, the forcing-scale resolving result can be applied to analyze measured data to obtain information on external forcing.

The structure of this paper is as follows. We present our derivation in \S \ref{sec_dev}, in which the details of the third- and fifth-order structure functions are shown in \S \ref{sec_du3} and \ref{sec_du5}, respectively, and the procedure for calculating the general odd high-order structure functions is shown in \S \ref{sec_high_order}. We numerically check our theoretical results in \S \ref{sec_num}. Finally, we summarize and discuss our results in \S \ref{sec_sum}.

\section{Derivation of the global expression for structure functions}\label{sec_dev}

We study the one-dimensional Burgers equation
\begin{equation}
	u_t + uu_x = F + D, \label{Burgers0}
\end{equation}
where $F$ and $D$ are external forcing and dissipation terms to be specified.

One special property for Burgers equation is that when the forcing and dissipation effects are absent, i.e., $F=0$ and $D=0$, all high-order momentums are preserved:
\begin{equation}
	\frac{1}{n}u^n_t + \frac{1}{n+1}\pa{x}u^{n+1} = 0,
\end{equation}
where $n\neq-1$ is a arbitrary number.
This flux form is important to derive high-order structure-function relations \cite{Falkovich2010}.

In a statistically steady turbulent state, we consider two-point measurements at locations $x$ and $x'=x+r$ with the displacement $r$. By assuming homogeneity we obtain 
\begin{equation}
	\pa{x} = -\pa{x'} = -\pa{r}. \label{deri}
\end{equation}

Multiplying the governing equation at one point, $x$, with velocity at the other point, $u'=u(x')$, adding the conjugate equation, and taking the ensemble average, we obtain equations for correlations with arbitrary $a$ and $b$ ($a\neq1$ and $b\neq1$)
\begin{equation}\label{khm_gen}
	\begin{aligned}
		&\pa{t}\ovl{u'^au^b} + \pa{r} \br{ \frac{a}{a+1}\ovl{u'^{a+1}u^b} - \frac{b}{b+1}\ovl{u'^au^{b+1}} } \\= &a\ovl{u'^{a-1}u^bF'} + b\ovl{u'^au^{b-1}F} + a\ovl{u'^{a-1}u^bD'} + b\ovl{u'^au^{b-1}D},
	\end{aligned}
\end{equation} 
where the overbar denotes the ensemble average and the homogeneity relation (\ref{deri}) is used.

In one dimension, high-order structure functions are simply linked to correlation functions as follow
\begin{subequations}\label{du_exp}
	\begin{align}
		\ovl{\delta u^3} &= -3\ovl{u'^2u} + 3\ovl{u'u^2}, \label{du_exp_3}\\
		\ovl{\delta u^4} &= \ovl{u'^4} -4\ovl{u'^3u} + 6\ovl{u'^2u^2} - 4\ovl{u'u^3} + \ovl{u'^4},\\
		\ovl{\delta u^5} &= -5\ovl{u'^4u} + 10\ovl{u'^3u^2} -10\ovl{u'^2u^3} + 5\ovl{u'u^4}, \label{du_5}\\
		... &= ....
	\end{align}
\end{subequations}

Our present work based on the idea that we can use linear combination of the advection terms (the second term on the l.h.s.) in (\ref{khm_gen}) to recover the r.h.s. of (\ref{du_exp}) for the odd-order, $2n+1$-order with $n$ a integer, structure functions.
Moreover, due to the quadratic nonlinearity, we need to use the linear combination of the equations (\ref{khm_gen}) with $a+b=2n$.
The coefficients of the linear combination are obtained by solving a linear equation system with $n$ equations and $n$ variables. 
Unfortunately, this is not possible for even-power, say $2m$th-order structure functions, because the number of equations, $m$, is larger than the number of variables, $m-1$.


When the external forcing is white-noise in time with zero mean,  we can calculate the forcing impact:
\begin{equation}
	\begin{aligned}
		\dd\br{u'^au^b} =& \br{u'+\dd u'}^a\br{u+\dd u}^b-{u'^au^b}&\\
		=& au'^{a-1}u^b\dd u' + bu'^au^{b-1}\dd u  &\\
		&+ \frac{a(a-1)}{2}u'^{a-2}u^b\dd u'^2 + \frac{b(b-1)}{2}u'^au^{b-2}\dd u^2 + abu'^{a-1}u^{b-1}\dd u'\dd u + h.o.t..&\\
	\end{aligned}
\end{equation}
Since $\dd u = \dd F + O(\dd t)$, taking the ensemble average and to the leading order we obtain
\begin{equation}\label{F_wn}
	\begin{aligned}
		\frac{\dd{\ovl{u'^au^b}}}{\dd t} = \frac{a(a-1)}{2}\ovl{u'^{a-2}u^b} C(0) + \frac{b(b-1)}{2}\ovl{u'^au^{b-2}}C(0) + ab\ovl{u'^{a-1}u^{b-1}}C(r) + h.o.t..,\\
	\end{aligned}
\end{equation}
where the terms linear in $\dd u$ or $\dd u'$ are identical to zero due to the external forcing's zero mean.

The white-noise forcing implies that
\begin{equation}
	\ovl{\dd F' \dd F} = C(r)\dd t
\end{equation}
with $C(r)$ is the externally prescribed forcing correlation.
And when defining
\begin{equation}
	P = \frac{C}{2} = \frac{1}{2}\br{\ovl{uF'}+\ovl{u'F}}
\end{equation}
we have $P(0)=\epsilon$, the energy injection rate.

Thus, we can write (\ref{khm_gen}) as
\begin{equation}\label{khm_gen_white}
	\begin{aligned}
		&\ovl{u'^au^b}_t + \pa{x} \br{ \frac{a}{a+1}\ovl{u'^{a+1}u^b} - \frac{b}{b+1}\ovl{u'^au^{b+1}} } \\
		&= \frac{a(a-1)}{2}\ovl{u'^{a-2}u^b} C(0) + \frac{b(b-1)}{2}\ovl{u'^au^{b-2}}C(0) + ab\ovl{u'^{a-1}u^{b-1}}C(r) + a\ovl{u'^{a-1}u^bD'} + b\ovl{u'^au^{b-1}D}.
	\end{aligned}
\end{equation} 
Here, we have not prescribe the form of damping since other than the linear damping most damping forms bring no simplification. 
And we will justify that in the limit of infinite Reynolds number the impact of dissipation on the structure functions vanishes.
Note that differing from the derivation for the inertial range results \cite{Falkovich2011} where the forcing effect is taken to be a constant, we calculate the globally valid expression for the forcing effect, which later makes our results for structure functions valid beyond the inertial range.  Therefore our structure-function expressions are more easily checked by numerical simulations since identifying the inertial ranges of numerical data may brings about artificial error.

We present the details of the third- and fifth-order structure functions and sketch the procedure of calculating other high-order structure functions in the following subsections. 
The expression for the third-order structure resolves the forcing scale similar to those obtained in \cite{Xie2018,Xie2019,Xie2019b} for 2D turbulence, while the expression for the fifth and high-order structure functions are special new features for the one-dimensional Burgers turbulence.

\subsection{Exact relation for $\ovl{\delta u^3}$}     \label{sec_du3}
Choosing $a=b=1$ in (\ref{khm_gen}) we obtain
\begin{equation}\label{khm_3}
	\pa{t}\ovl{u'u} + \pa{r} \br{ \frac{1}{2}\ovl{u'^{2}u} - \frac{1}{2}\ovl{u'u^{2}} } = 2P + \ovl{uD'} + \ovl{u'D}.
\end{equation} 
In a statistically steady state, using (\ref{du_exp_3}), (\ref{khm_3}) becomes
\begin{equation}\label{khm_3_steady}
	-\frac{1}{6}\pa{r} \ovl{\delta u^3} = 2P + \ovl{uD'} + \ovl{u'D}.
\end{equation}
When the external forcing is white-noise in time, using (\ref{khm_gen_white}) and taking $D=0$ we obtain
\begin{equation}
	\ovl{\delta u^3} = -6 \int_{0}^{r} C(s) \dd s, \label{exact_du3}
\end{equation}
which is a forcing-scale-resolving exact result (cf. \cite{Bec2000}).

Here, it is legal to take $D=0$ since energy transfers downscale in the Burgers turbulence. We can justify this argument based on dimensional analysis which implies that the contribution of small-scale viscosity tends to zero as the viscosity tends to zero (cf. \cite{Xie2018} for 2D turbulence).

In the limit of small $r$ we obtain 
\begin{equation}
	\ovl{\delta u^3} = -12 \epsilon r,
\end{equation}
which recovers the classic inertial-range results \cite{E1999}.

Note that when the external forcing centres at one wavenumber the expression (\ref{exact_du3}) is already more generally calculated in \cite{Xie2019} with bidirectional energy transfer, but the coefficient $4$ in their expression (4.1) should be replaced by $12$.

\subsection{Exact relation for $\ovl{\delta u^5}$}    \label{sec_du5}

Because of the quadratic nonlinearity in the Burger equation (\ref{Burgers0}), to obtain the fifth-order structure function we consider the correlation equations with $a+b=4$ (cf. (\ref{khm_gen})):
\begin{subequations}\label{khm_5}
	\begin{align}
		\ovl{u'^3u}_t + \pa{r} \br{ \frac{3}{4}\ovl{u'^{4}u} - \frac{1}{2}\ovl{u'^3u^{2}} } &= 3\ovl{u'^{2}uF'} + \ovl{u'^3F} + 3\ovl{u'^{2}uD'} + \ovl{u'^3D}, \label{khm_31}\\
		\ovl{u'^2u^2}_t + \pa{r} \br{ \frac{2}{3}\ovl{u'^{3}u^2} - \frac{2}{3}\ovl{u'^2u^{3}} } &= 2\ovl{u'u^2F'} + 2\ovl{u'^2uF} + 2\ovl{u'u^2D'} + 2\ovl{u'^2uD}, \label{khm_22}\\
		\ovl{u'u^3}_t + \pa{r} \br{ \frac{1}{2}\ovl{u'^{2}u^3} - \frac{3}{4}\ovl{u'u^{4}} } &= \ovl{u^3F'} + 3\ovl{u'u^{2}F} + \ovl{u^3D'} + 3\ovl{u'u^{2}D}. \label{khm_13}
	\end{align}
\end{subequations}

To get the expression for $\ovl{\delta u^5}$ 
(cf. (\ref{du_5})), 
we take the combination: $-\dfrac{20}{3}$(\ref{khm_31}) $+10$(\ref{khm_22})$ -\dfrac{20}{3}$(\ref{khm_13}), which in a statistically state leads to
\begin{equation}\label{exact_du5_0}
	\begin{aligned}
		\pa{r} \ovl{\delta u^5} =& -20 \ovl{u'^{2}uF'}  - \frac{20}{3}\ovl{u'^3F} + 20\ovl{u'u^2F'} + 20\ovl{u'^2uF} -\frac{20}{3}\ovl{u^3F'} - 20\ovl{u'u^{2}F}\\
		&-20 \ovl{u'^{2}uD'}  - \frac{20}{3}\ovl{u'^3D} + 20\ovl{u'u^2D'} + 20\ovl{u'^2uD} -\frac{20}{3}\ovl{u^3D'} - 20\ovl{u'u^{2}D}.
	\end{aligned}
\end{equation}

When the external forcing is white-noise in time, by making use of (\ref{khm_gen_white}) and taking $D=0$ we obtain
\begin{equation}\label{exact_5}
	\pa{r} \ovl{\delta u^5} = 20\br{ (C(0)-C(r))\ovl{\delta u^2} - C(0)\ovl{u^2} },
\end{equation}
which is an exact global relation linking the 2nd- and 5th-order structure functions in the ``zero-viscosity limit" in the range away from the viscous scale.
{Here, for high-order conserved quantities $E_n = \int u^{2n} \dd \bx/2$, corresponding to the KHM equations with $a+b=2n$, the viscous scale can be defined as $l_{\nu,n} = \nu^{(2n+1)/(2n+2)}\epsilon_n^{-(2n+2)}$ with $\epsilon_n$ defined as the dissipation rate for $E_n$.
	Note that $l_{\nu,1}$ is the Kolmogorov viscous scale.
	Our zero-viscosity limiting process bases on the fact that the conserved quantities transfer downscale and finally dissipate at small scales, therefore in the corresponding KHM  equation, the dissipation effect is only of $O(1)$ in the viscous range $r\leq l_{\nu,n}$, while when the scale is much larger than the viscous scale $l_{\nu,n}$ the dominate balance is between the nonlinear term and the forcing term. 
	We justify our limiting process by the numerical simulations in the next section.}

In the sublimit $r\to 0$, (\ref{exact_5}) implies the inertial range result
\begin{equation}
	\ovl{\delta u^5} = -20 \ovl{u^2}C(0) r,
\end{equation}
which consists with (16) in \cite{E1999}. 

\subsection{Higher-odd-order exact relations}\label{sec_high_order}

Similar to the calculation of $\ovl{\delta u^5}$, considering
\begin{equation}
	\delta u^{2n+1} = -C_{2n+1}^1 u'^{2n}u + ... + (-1)^j C_{2n+1}^j u'^{2n+1-j}u^j + ... + C_{2n+1}^{2n}u'u^{2n}, \label{du_2n1}
\end{equation}
we make linear combination of (\ref{khm_gen}) with $a+b=2n$ to calculate the expressions for the odd high-order structure functions. 
This is possible due to the symmetry when $a+b=2n$ and therefore we only need to solve $n$ variables from $n$ equations. 

In this linear combination, we define the coefficients in front of the equations $(a=2n-j,b=j)$ and $(a=j,b=2n-j)$ as $d_j$ due to symmetry. Then, (\ref{du_2n1}) and (\ref{khm_gen}) lead to the following equation for $d_j$:
\begin{equation}\label{d_eq}
	\begin{bmatrix}
		C_{2n+1}^1 \\ -C_{2n+1}^2 \\ ... \\ (-1)^{j+1}C_{2n+1}^j \\ ... \\ (-1)^{n+1}C_{2n+1}^n
	\end{bmatrix}
	=
	\underset{M}{\underbrace{\begin{bmatrix}
				-\frac{2n-1}{2n} & & & & &\\
				\frac{1}{2} & - \frac{2n-2}{2n-1} & & & &\\
				&  & ... & & &\\
				&  &  \frac{j-1}{j} & - \frac{2n-j}{2n-j+1} & &\\
				&  & & & ... &\\
				&  & & & \frac{n-1}{n} & -\frac{n}{n+1}\\
	\end{bmatrix}}}
	\underset{\boldsymbol{d}}{\underbrace{\begin{bmatrix}
				d_1 \\ d_2 \\ ... \\ d_j \\ ... \\ d_n 
	\end{bmatrix}}},
\end{equation}
where the lower-triangular coefficient matrix $M$ enables us to solve it by back substitution. 

After solving $\boldsymbol{d}$, in a statistically steady state, taking $D=0$ and assuming a temporal white-noise forcing (cf. (\ref{khm_gen_white})),  we obtain
\begin{equation}\label{SF_2n+1}
	\pa{r}\ovl{\delta u^{2n+1}} = D_1(\ovl{u'^au^b},\boldsymbol{d}) C(0) + D_2(\ovl{u'^au^b},\boldsymbol{d}) C(r),
\end{equation}
where $D_i$ depend on the correlation functions with $a+b=2n-2$.

In the limit of small $r$, (\ref{SF_2n+1}) reduces to the inertial-range result
\begin{equation}
	\ovl{\delta u^{2n+1}} = D(\boldsymbol{d}) \ovl{u^{2n-2}}C(0) r, \label{du_2n+1_limit}
\end{equation}
where $D(\boldsymbol{d}) = n(2n-1)\br{2\sum_{j=1}^{n-1}d_j+d_n}=-2n(2n+1)<0$.
The negative signs confirm that the high-order conserved quantities transfer downscale, which is consistent with taking $D=0$.
The limiting result (\ref{du_2n+1_limit}) again is consistent with the results of \cite{E1999} which is calculated based on the statistics of shocks.
{In addition, considering that $2\epsilon_n = n(2n-1)\ovl{u^{2n-2}}C(0)$, we can express (\ref{du_2n+1_limit}) as
	\begin{equation}
		\ovl{\delta u^{2n+1}} = - \frac{4(2n-1)}{2n+1}\epsilon_n r,
	\end{equation}
	which recovers the result obtained by \cite{Falkovich2006} and \cite{Cardy2008}.
	
	\section{Numerical simulations} \label{sec_num}
	
	In this section, we run second-order finite-volume numerical simulations for the temporal white-noise forced Burgers equation to check the above derived theoretical results. The resolution varies with a periodic domain size of $2\pi$. 
	We add no explicit dissipation and the injected energy is absorbed by the numerical dissipation of the finite-volume scheme to reach statistically steady states. 
	
	We specify the external forcing as
	\begin{equation}
		F = A_1\sin\br{k_1x} + B_1\cos\br{k_1x} + A_2\sin\br{k_2x} + B_2\cos\br{k_2x},
	\end{equation} 
	where $A_i$ and $B_i$ are Gaussian Random variables with variation $\sigma^2=0.04$, and we choose $k_1=3$ and $k_2=17$ to fully test our globally valid theoretical result.
	Thus, we can analytically calculate the forcing correlation function
	\begin{equation}
		C(r) = 2\sigma^2 \br{\cos{k_1 r}+ \cos{k_2 r}}.
	\end{equation}

	The expression of third-order structure function (\ref{exact_du3}) is checked in Figure \ref{fig_du3}.
	The left panel of Figure \ref{fig_du3} shows that the numerical and theoretical results match well not only in the short inertial range but also globally.
	And our global expression matches well the inertial-range result in the limit of small displacement.
	The discrepancy brings about by the numerical dissipation is localized around $r=0$ as what is shown in the right panel of Figure \ref{fig_du3}.
	\begin{figure}
		\centering
		\includegraphics[width=0.49\linewidth]{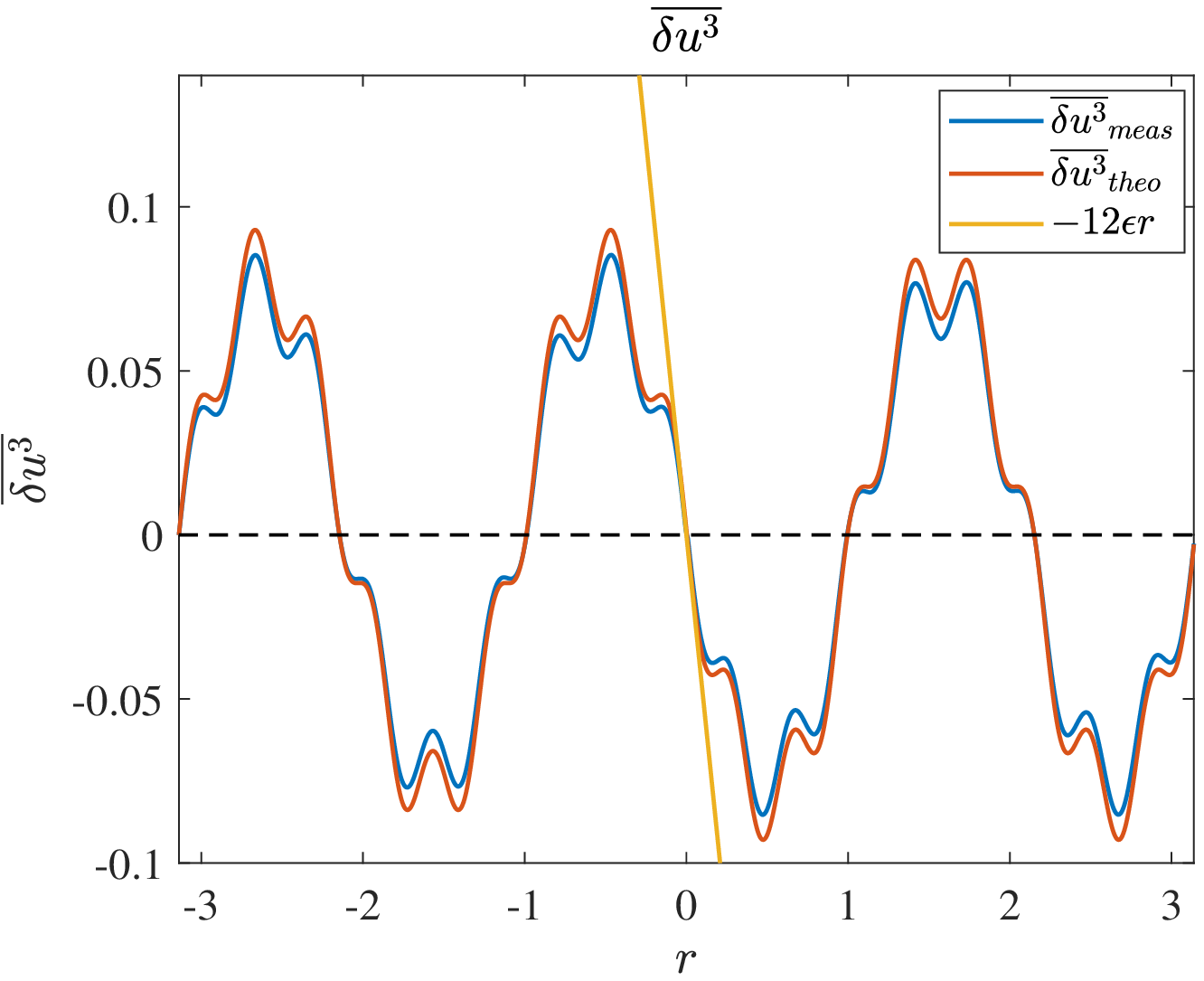}
		\includegraphics[width=0.49\linewidth]{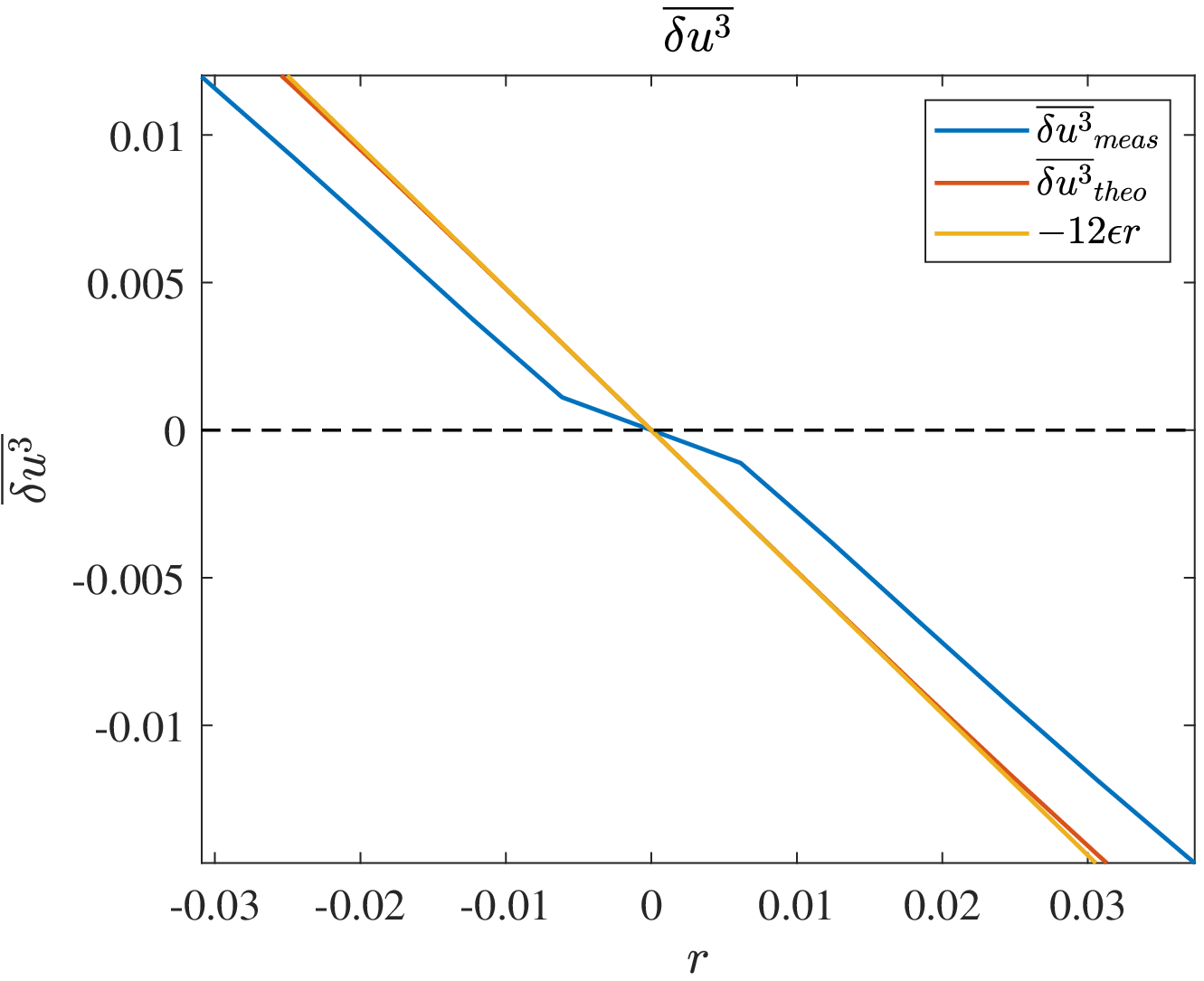}
		\caption{Comparison between the numerically obtained and theoretical expression for third-order structure function. The right panel zooms in the small $r$ region of the left panel.}
		\label{fig_du3}
	\end{figure}
	
	We show the second-order structure function in the left panel of Figure \ref{fig_du2_5}.
	Then we substitute this numerical result into the r.h.s. of (\ref{exact_5}) to calculate the fifth-order structure function. 
	Based on the expression of (\ref{exact_5}), the domain average of both sides should be zero because the l.h.s. has a derivative form in a periodic domain, however, after substituting the numerically obtained second-order structure function we find the domain average of the r.h.s. is nonzero. 
	So in the theoretical result of the fifth-order structure function shown in the right panel of Figure \ref{fig_du2_5} we subtracted the domain-averaged mean value in the expression of $\ovl{\delta u^5}_r$:
	\begin{equation}
		\pa{r}\ovl{\delta u^5}_{theo} = 20\br{ (C(0)-C(r))\ovl{\delta u^2} - C(0)\ovl{u^2} } - \av{20\br{ (C(0)-C(r))\ovl{\delta u^2} - C(0)\ovl{u^2} }}
	\end{equation}
	where the angle bracket denoting the domain average. And correspondingly in the small $r$ limit we define 
	\begin{equation}
		\ovl{u^2}C(0)^* = \ovl{u^2}C(0) - \av{\br{ (C(0)-C(r))\ovl{\delta u^2} - C(0)\ovl{u^2} }}.
	\end{equation}
	\begin{figure}
		\centering
		\includegraphics[width=0.49\linewidth]{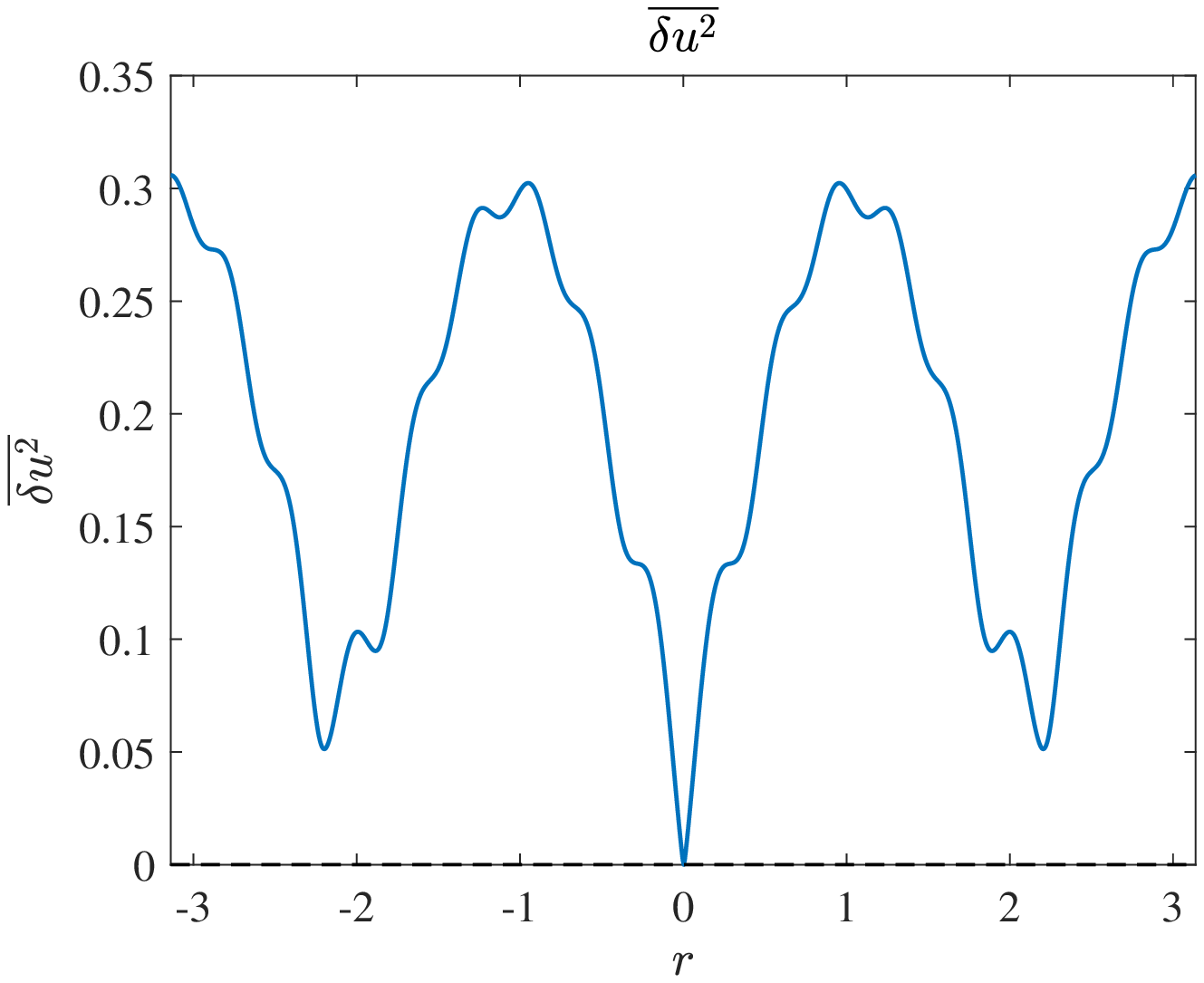}
		\includegraphics[width=0.49\linewidth]{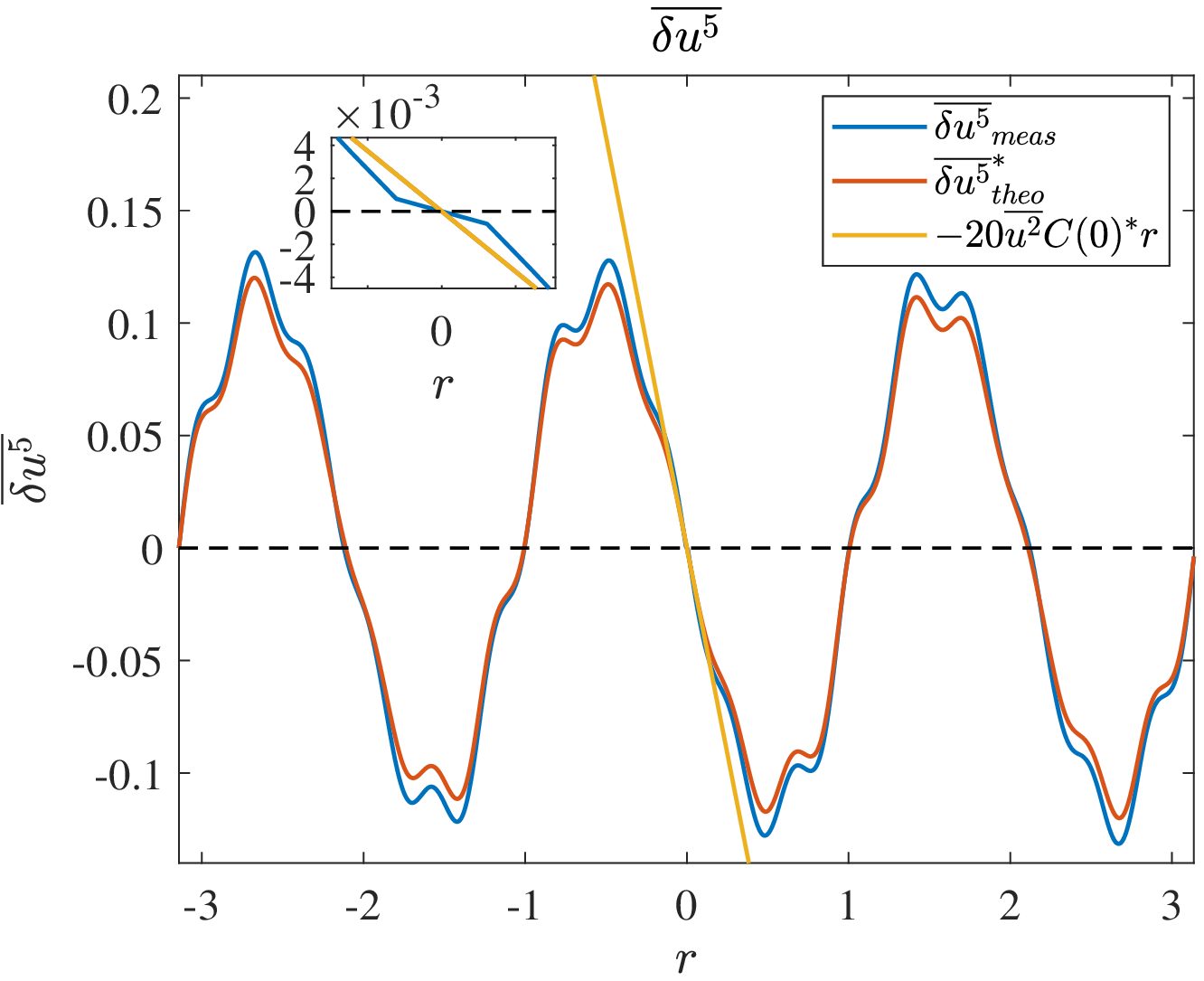}
		\caption{Numerically obtained second- and fifth-order structure functions.}
		\label{fig_du2_5}
	\end{figure}
	
	After subtracting the domain average, we find in the right panel of Figure \ref{fig_du2_5} that the theoretical and numerical results match well globally and the classic inertial-range result is recovered. 
	Also, the inset in the right panel of Figure \ref{fig_du2_5} we show that similar to the impact on the third-order structure function, the numerical and theoretical result of the fifth-order structure function has discrepancy around $r=0$ due to the numerical dissipation.
	
	
	The numerical dissipation is also the reason for the domain-average discrepancy in (\ref{exact_5}).
	The finite-volume scheme captures the weak solution to the viscous Burgers equation with dissipation term $D= \nu\nabla^2u$ in (\ref{Burgers0}) in the limit $\nu \to 0$, but the finite-resolution numerical simulation makes the finite-$\nu$ effect manifests. 
	Therefore in \S \ref{sec_5th_vis} we calculate the normal viscosity's impact on the fifth-order structure function, $D_5$. 
	We justify in Figure \ref{fig_vis5} that the domain-average discrepancy is a result of numerical dissipation by the matching of the theoretical and numerical viscous effects.
	Since we do not know the value of $\nu$, in Figure \ref{fig_vis5} the numerical viscosity is fitted as $\nu=7\times10^{-4}$.
	\begin{figure}
		\centering
		\includegraphics[width=0.49\linewidth]{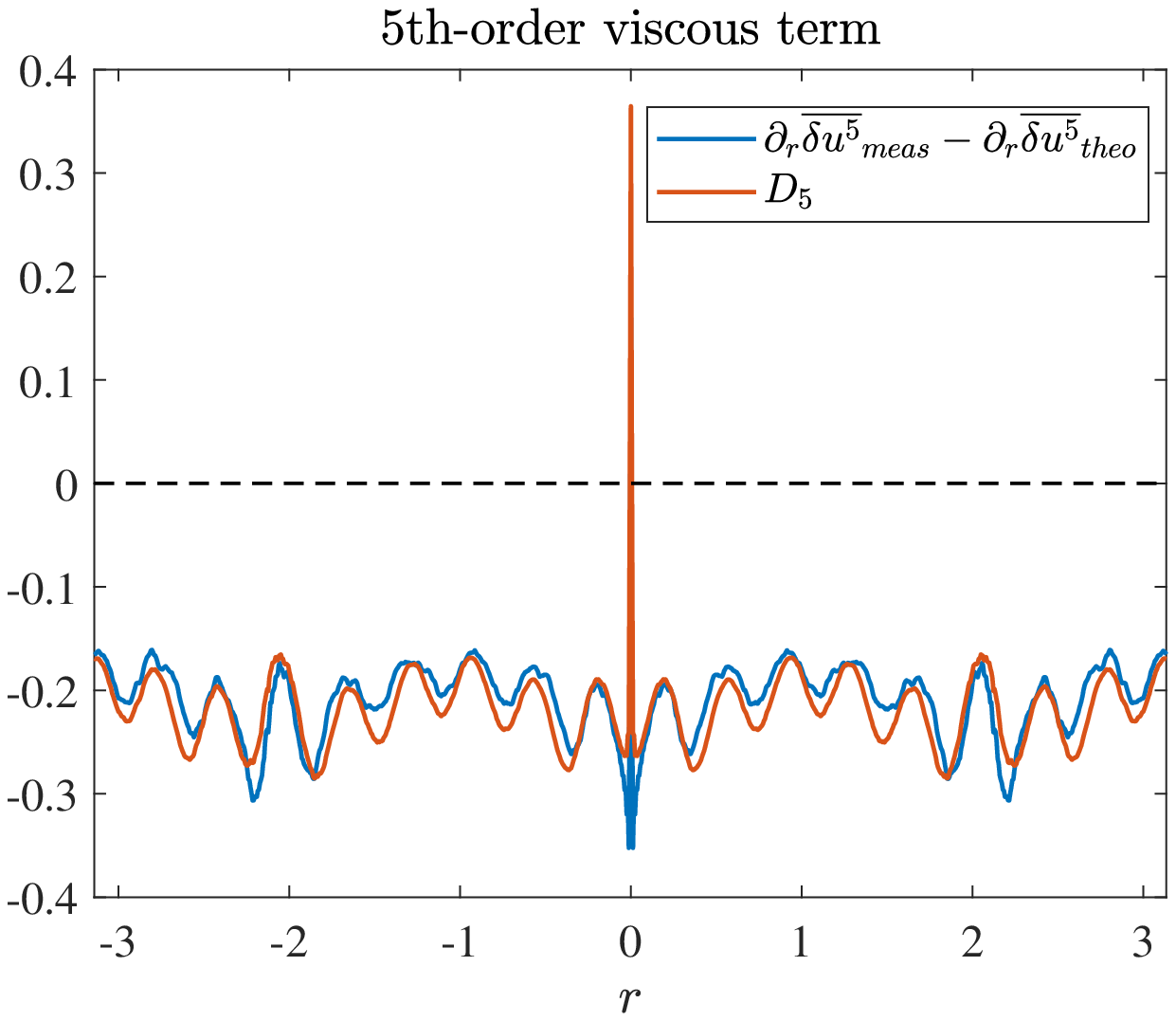}
		\caption{Comparison of the $\pa{r} \ovl{\delta u^5}-\pa{r} \ovl{\delta u^5}_{theo}$ and the dissipation term $D_5$ (cf. \S \ref{sec_5th_vis}).}
		\label{fig_vis5}
	\end{figure}
	
	Our theoretical derivation where $D=0$ is taken requires the justification of a vanishing viscous impact in the limit of infinite resolution. 
	So we run numerical simulations with resolutions $512,\,1024,\,2048$ and $4096$ to obtain the best-fit viscosities.
	In Figure \ref{fig_res_nu}, we find that the numerical viscosity is inversely proportional to the resolution, indicating a vanishing dissipation effect in the limit of infinite resolution. This inverse proportion is understood as the first-order accuracy of the finite-volume scheme in capturing the shocks. This justifies our theoretical procedure of taking $D=0$ to study the zero viscosity limit.
	\begin{figure}
		\centering
		\includegraphics[width=0.49\linewidth]{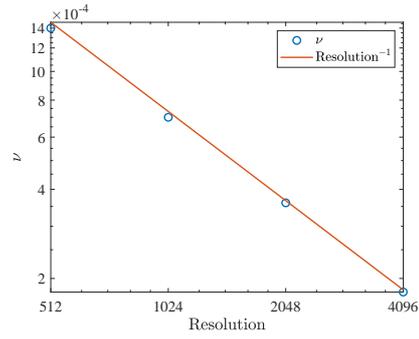}
		\caption{Dependence of the numerical viscosity on resolution.}
		\label{fig_res_nu}
	\end{figure}
	

	\section{Summary and discussion} \label{sec_sum}
	
	By studying high-order KHM equations, we obtain forcing-scale-resolving exact relations between the $\br{2n+1}$th and $\br{2n-2}$th-order structure functions for Burgers turbulence with temporal white-noise forcing in the limit of zero viscosity. 
	In the sublimit of the small distance of two measured points, our result recovers the classic linear dependence of the high-order structure function on the distance (cf. \cite{E1999,Falkovich2006,Falkovich2011}).   
	In the special case of $n=1$, Kolmogorov-like result is recovered where the third-order structure function is proportional to the energy injection rage and the distance of two measured points.
	Compared with \cite{Falkovich2006} and \cite{Cardy2008}, we take the limit of zero viscosity to the very last step, which allows us to obtain global relations for structure beyond the inertial range. 
	
	We check our results using finite-volume numerical simulations, where finite numerical dissipation must present.
	The viscous effect only impacts the third-order structure function expression at $r=0$, but for higher-order structure functions, the viscosity introduces finite domain-averaged discrepancy.
	By tuning the resolution, we show that the viscous effect tends to zero as the resolution increases, therefore we justify our results' validity in the limit of zero viscosity (infinite resolution).
	This check of the dissipation effect is in consistent with the existence of dissipation anomalies in compressible turbulence \cite{Eyink2018}.
	
	Comparing with the results in \cite{E1999}, our result is also valid in the forcing scale besides from the inertial range; also in our expression, the structure function are the only quantities that are required, while their results need the statistics of the shocks which are harder to obtain and therefore not easy to be justified by numerical simulations. 
	But the shock physics is important for our calculation. The conserved quantities dissipate dominantly at the shocks, which has a characteristic scale proportional to the viscosity. So in the limit of zero viscosity, the effect of dissipation is confined at small scales, leaving the dominant balance between the nonlinear term and the forcing term in the KHM equations as long as the scale considered is much larger than the viscous scale.
	Nevertheless, our calculation is limited to odd-order expressions.
	
	We show that the KHM equation and its high-order forms are suitable to study the structure functions in the zero viscosity limit by merely taking the viscosity equal to zero due to the downscale energy flux. 
	We have reasoned this procedure in \cite{Xie2018} for the third-order structure functions in two-dimensional vortex turbulence. 
	In this paper, by studying the Burgers equation we show that this procedure of dealing with the zero viscosity limit is also valid for high orders. 
	Compared with the inertial-range expressions for odd-order structure functions which are also derived using the KHM approach \cite{Falkovich2011}, by resolving the forcing scale, our expressions are valid beyond the inertial range and therefore more easily justified by numerical simulations. 
	
	Even though the results obtained from one-dimensional Burgers turbulence is limited to its simple setup, when shocks are abundant the properties of two- and three-dimensional polytropic gas turbulence \cite{Augier2019,Lindborg2019} and three-dimensional compressible turbulence \cite{Wang2012} share similar properties with the one-dimensional Burgers turbulence. Our future work will seek the implication of our high-order structure-function results on high-dimensional compressible turbulence.
	Considering that different types of external forcing lead to distinctive statistics in compressible turbulence (cf. \cite{Peterson2010,Jagannathan2016}), the global expressions that link forcing statistics with structure function would be practically useful because they can obtain external forcing information using statistics of measured data .\\ 

\acknowledgement{J-HX thanks Wei Chen for an elegant analytical method to solve (20).
	J-HX greatfully acknowledges financial support from the National Natural Science foundation of China grant No. 92052102.}

\appendix

\section{Viscous impact on the fifth-order structure function}\label{sec_5th_vis}

In this section we consider Burgers equation with normal viscosity:
\begin{equation}
	u_t + uu_x = F + \nu u_{xx},
\end{equation}
and focus on the viscous impact on the fifth-order structure function.

Substituting $D=\nu\nabla^2u$ into (\ref{exact_du5_0}) and assuming the temporal white-noise forcing, we obtain
\begin{equation}
	\pa{r} \ovl{\delta u^5} = D_5 + \underset{\pa{r} \ovl{\delta u^5}_{theo}}{\underbrace{20\br{ (C(0)-C(r))\ovl{\delta u^2} - C(0)\ovl{u^2} }}},
\end{equation}
with
\begin{equation}
	D_5 = \nu \lbr{ \pa{rr}\sbr{ -\frac{40}{3}\br{\ovl{u'^3u}+\ovl{u'u^3}} + 20\ovl{u'^2u^2} } -20 \ovl{ \br{{u_x}^2+{u'_{x'}}^2 }\delta u^2 }  }.
\end{equation}
Comparing with (\ref{exact_5}), the dissipation effect is captured by $D_5$.

\bibliographystyle{acm}
\bibliography{turb.bib}

\end{document}